# Approximating the Online Set Multicover Problems Via Randomized Winnowing[*]

Piotr Berman[†]
Department of Computer Science and Engineering
Pennsylvania State University
University Park, PA 16802
Email: berman@cse.psu.edu

Bhaskar DasGupta[‡]
Department of Computer Science
University of Illinois at Chicago
Chicago, IL 60607
Email: dasgupta@cs.uic.edu

August 27, 2018

**Abstract**

In this paper, we consider the weighted online set $k$-multicover problem. In this problem, we have a universe $V$ of elements, a family $\mathcal{S}$ of subsets of $V$ with a positive real cost for every $S \in \mathcal{S}$, and a "coverage factor" (positive integer) $k$. A subset $\{i_0, i_1, \ldots\} \subseteq V$ of elements are presented online in an arbitrary order. When each element $i_p$ is presented, we are also told the collection of all (at least $k$) sets $\mathcal{S}_{i_p} \subseteq \mathcal{S}$ and their costs to which $i_p$ belongs and we need to select additional sets from $\mathcal{S}_{i_p}$ if necessary such that our collection of selected sets contains *at least* $k$ sets that contain the element $i_p$. The goal is to *minimize* the *total cost* of the selected sets[1]. In this paper, we describe a new randomized algorithm for the online multicover problem based on a randomized version of the winnowing approach of [15]. This algorithm generalizes and improves some earlier results in [1, 2]. We also discuss lower bounds on competitive ratios for *deterministic algorithms* for general $k$ based on the approaches in [2].

## 1 Introduction

In this paper, we consider the Weighted Online Set $k$-multicover problem (abbreviated as **WOSC**$_k$) defined as follows. We have an universe $V = \{1, 2, \ldots, n\}$ of elements, a family $\mathcal{S}$ of subsets of $U$ with a cost (positive real number) $c_S$ for every $S \in \mathcal{S}$, and a "coverage factor" (positive integer) $k$. A subset $\{i_0, i_1, \ldots\} \subseteq V$ of elements are presented in an arbitrary order. When each element $i_p$ is presented, we are also told the collection of all (at least $k$) sets $\mathcal{S}_{i_p} \subseteq \mathcal{S}$ in which $i_p$ belongs and we need to select additional sets from $\mathcal{S}_{i_p}$, if necessary, such that our collection of sets contains *at least* $k$ sets that contain the element $i_p$. The goal is to minimize the total cost of the selected sets. The special case of $k = 1$ will be simply denoted by **WOSC** (Weighted Online Set Cover). The unweighted versions of these problems, when the cost any set is one, will be denoted by **OSC**$_k$ or **OSC**.

---

[*]A preliminary version of these results appeared in $9^{\text{th}}$ Workshop on Algorithms and Data Structures, F. Dehne, A. López-Ortiz and J. R. Sack (editors), LNCS 3608, pp. 110-121, 2005.
[†]Supported by NSF grant CCR-0208821.
[‡]Supported in part by NSF grants DBI-0543365, IIS-0612044 and IIS-0346973.

[1]Our algorithm and competitive ratio bounds can be extended to the case when a set can be selected at most a prespecified number of times instead of just once; we do not report these extensions for simplicity and also because they have no relevance to the biological applications that motivated our work.

The performance of any online algorithm can be measured by the *competitive ratio*, *i.e.*, the ratio of the total cost of the online algorithm to that of an optimal offline algorithm that knows the entire input in advance; for randomized algorithms, we measure the performance by the *expected* competitive ratio, *i.e.*, the ratio of the expected cost of the solution found by our algorithm to the optimum cost computed by an adversary that knows the entire input sequence and has no limits on computational power, but who is *not familiar* with our random choices.

The following notations will be used uniformly throughout the rest of the paper unless otherwise stated explicitly:

- $V$ is the universe of elements;

- $m = \max_{i \in V} |\{S \in \mathcal{S} \mid i \in S\}|$ is the maximum *frequency*, *i.e.*, the maximum number of sets in which any element of $V$ belongs;

- $d = \max_{S \in \mathcal{S}} |S|$ is the maximum set size;

- $k$ is the coverage factor;

- $e$ is the base of natural logarithm.

*None of $m$, $d$ or $|V|$ is known to the online algorithm in advance.*

## 1.1 Motivations and applications

One of our main motivation for investigating these problems, especially for large values of the "coverage factor", is their applications to reverse engineering problems in systems biology. However, other applications have also been noted in previous literatures and below we mention one such application in addition to the biological motivations.

### 1.1.1 Client/server protocols [2]

Such a situation is modeled by the problem **WOSC** in which there is a network of servers, clients arrive one-by-one in arbitrary order, and each client can be served by a subset of the servers based on their geographical distance from the client. The extension to **WOSC**$_k$ handles the scenario in which a client must be attended to by at least a minimum number of servers for, say, reliability, robustness and improved response time. In addition, in our motivation, we want a distributed algorithm for the various servers, namely an algorithm in which each server locally decide about the requests without communicating with the other servers or knowing their actions (and, thus for example, not allowed to maintain a potential function based on a subset of the servers such as in [2]).

### 1.1.2 Reverse engineering of gene/protein networks [4, 7, 8, 10, 13, 14, 18, 20]

We briefly explain this motivation here due to lack of space; the reader may consult the references for more details. This motivation concerns unraveling (or "reverse engineering") the web of interactions among the components of complex protein and genetic regulatory networks by observing global changes to derive interactions between individual nodes. In this application our attention is focused solely on one such approach, originally described in [13, 14], further elaborated upon in [4, 18], and reviewed in [10, 20]. Here one assumes that the time evolution of a vector of state variables $x(t) = (x_1(t), \ldots, x_n(t))$ is described by a system of differential equations:

$$\frac{\partial \vec{x}}{\partial t} = f(\vec{x}, \vec{p}) \equiv \begin{cases} \frac{\partial x_1}{\partial t} &= f_1(x_1, \ldots, x_n, p_1, \ldots, p_m) \\ \frac{\partial x_2}{\partial t} &= f_2(x_1, \ldots, x_n, p_1, \ldots, p_m) \\ &\vdots \\ \frac{\partial x_n}{\partial t} &= f_n(x_1, \ldots, x_n, p_1, \ldots, p_m) \end{cases}$$

where $\vec{p} = (p_1, \ldots, p_m)$ is a vector of parameters, such as levels of hormones or of enzymes, whose half-lives are long compared to the rate at which the variables evolve and which can be manipulated but remain constant during any given experiment. The components $x_i(t)$ of the state vector represent quantities that can be in principle measured, such as levels of activity of selected proteins or transcription rates of certain genes. There is a reference value $\bar{p}$ of $\vec{p}$, which represents "wild type" (that is, normal) conditions, and a corresponding steady state $\bar{x}$ of $\vec{x}$, such that $f(\bar{x}, \bar{p}) = 0$. We are interested in obtaining information about the Jacobian of the vector field $f$ evaluated at $(\bar{x}, \bar{p})$, or at least about the signs of the derivatives $\partial f_i/\partial x_j(\bar{x}, \bar{p})$. For example, if $\partial f_i/\partial x_j > 0$, this means that $x_j$ has a positive (catalytic) effect upon the rate of formation of $x_i$. To be more precise, the goal is to find as much information as possible about an unknown matrix $A \in \mathbb{R}^{n \times n}$ which is the Jacobian matrix $\partial f/\partial x$. The critical assumption is that, while we may not know the form of $f$, we often do know that *certain parameters $p_j$ do not directly affect certain variables $x_i$*. This amounts to *a priori* biological knowledge of specificity of enzymes and similar data. Such a knowledge can be summarized by a binary matrix $C = (c_{ij}) \in \{0, 1\}^{n \times m}$, where "$c_{ij} = 1$" means that $p_j$ *does not* appear in the equation for $\dot{x}_i$, that is, $\partial f_i/\partial p_j \equiv 0$. In our current context, each row of $C$ correspond to an element, each column of $C$ correspond to a set, and 0-1 entries indicate the memberships of elements in sets. A crucial contribution of the above-mentioned references in this context is as follows. Suppose that we solve this set-multicover instance in which each element is covered at least some $\beta$ times. Then with $\beta = n - 1$ we can recover the elements of $A$ *uniquely up to a scalar multiple* (and, thus can know the signs of the derivatives $\partial f_i/\partial x_j(\bar{x}, \bar{p})$ precisely) and with $\beta = n - k$ for some small $k$ we can recover the elements of $A$ up to a modest ambiguity that can be tolerated in practice. If the corresponding experimental protocols are carried out using measurements via a suitable biological reporting mechanisms such as fluorescent proteins in an online fashion, one arrives at the online set multicover problems discussed in this paper.

## 1.2 Summary of prior work

Offline versions $\mathbf{WSC}_k$ and $\mathbf{SC}_k$ of the problems $\mathbf{WOSC}_k$ and $\mathbf{OSC}_k$, in which all the $|V|$ elements are presented at the same time, have been well studied in the literature. Following a brief summary of some of the results only about these problems. Assuming $\text{NP} \not\subseteq \text{DTIME}(n^{\log \log n})$, the $\mathbf{SC}_1$ problem in general cannot be approximated to within a factor of $(1 - \varepsilon) \ln |V|$ for any constant $0 < \varepsilon < 1$ in polynomial time [11] and cannot be approximated to within a factor of $\ln d - O(\ln \ln d)$ in polynomial time when restricted to set-cover instances with maximum set size $d$ for all sufficiently large $d$ unless P=NP. On the other hand, an instance of the $\mathbf{SC}_k$ problem can be $(1 + \ln d)$-approximated in $O(|V| \cdot |\mathcal{S}| \cdot k)$ time by a simple greedy heuristic that, at every step, selects a new set that covers the maximum number of those elements that has not been covered at least $k$ times yet [12, 22]; these results were recently improved upon in [7] who provided a randomized approximation algorithm with an expected performance ratio of $(1 + o(1)) \ln \left(\frac{d}{k}\right)$ when $d/k$ is at least about $e^2 \approx 7.39$, and for smaller values of $d/k$ the expected performance ratio was $1 + 2\sqrt{d/k}$.

Regarding previous results for the online versions, the authors in [1, 2] considered the **WOSC** problem and provided both deterministic and simple randomized algorithms with a competitive ratio or expected competitive ratio of $O(\log m \log |V|)$ and an almost matching lower bound of $\Omega \left( \frac{\log |\mathcal{S}| \log |V|}{\log \log |\mathcal{S}| + \log \log |V|} \right)$ on the competitive ratio for any deterministic algorithm for almost all values[2] of $|V|$ and $|\mathcal{S}|$. The authors of [5] provided an efficient randomized online approximation algorithm and a corresponding matching lower bound (for any randomized algorithm) for a different version of the online set-cover problem in which one is allowed to pick at most k sets for a given k and the goal is to maximize the number of presented elements for which at least one set containing them was selected on or before the element was presented. Concurrent to our conference publication, Alon, Azar and Gutner [3] considered the weighted online set-cover problem with repetitions which is studied in a bigger context of admissions control problem in general networks. Here, an element can be presented multiple times and, if the element is presented k times, our goal is to cover it by at least k different sets. For this problem [3] contains a randomized $O(\log m \log |V|)$-competitive algorithm as well as a deterministic bi-criteria approximation algorithm, *i.e.*, a deterministic algorithm that has a competitive ratio of $O(\log m \log |V|)$ and covers an element by at least $(1-\varepsilon)k$ different sets for any fixed $\varepsilon > 0$; it is easy to see that these bounds carry over to the problem **WOSC**$_k$. Conversely, it is not difficult to see that our algorithm **A-Universal** and analysis can easily be adapted to this problem to achieve an expected competitive ratio of $\log_2 m \ln d + O(\log_2 m + \ln d)$ with arbitrary set weights; one would need to modify appropriate places of Section 3.4. For unweighted sets, via Corollary 2(b), Algorithm **A-Universal** provides an improved expected competitive ratio of "roughly" (neglecting small constants) $\max \left\{ 5 \log_2 m, \log_2 m \ln \left( \frac{d}{k \log_2 m} \right) \right\}$ and the constants involved in this bound are further improved in Theorem 10.

### 1.3 Summary of our results and techniques

Let $r(m, d, k)$ denote the competitive ratio of any online algorithm for **WOSC**$_k$ as a function of $m$, $d$ and $k$. In this paper, we describe a new randomized algorithm for the online multicover problem based on a randomized version of the winnowing approach of [15]. Our main contributions are then as follows:

- We first provide a uniform analysis of our algorithm for all cases of the online set multicover problems. As a corollary of our analysis, we observe the following.

  – For **OSC**, **WOSC** and **WOSC**$_k$ our randomized algorithm has $\mathsf{E}[r(m, d, k)]$ equal to $\log_2 m \ln d$ plus small lower order terms. While the authors in [1, 2] did provide a deterministic algorithm and a simple randomized algorithm for **WOSC** with a competitive ratio and an expected competitive ratio of $O(\log m \log |V|)$, respectively, the improvements of our approach and analysis are as follows:

    * We provide better constant factors and lower-order terms. Note that tight analysis of the approximability or inapproximability bounds for set-cover type problems involving tight estimates of the constants and lower-order terms is not a new idea; for example, see [6, 7, 17, 19, 21].
    * We use the maximum set size d rather than the larger universe size $|V|$ in the competitive ratio bound.
    * For large coverage factor k (the case of utmost importance in our applications to systems biology in Section 1.1.2), our uniform analysis via the quantity $\kappa$ (see Section 3)

---

[2]To be precise, when $\log_2 |V| \leq |\mathcal{S}| \leq e^{|V|^{\frac{1}{2} - \delta}}$ for any fixed $\delta > 0$; we will refer to similar bounds as "almost all values" of these parameters in the sequel.

provides an expected competitive ratio of roughly

$$\max\left\{5\log_2 m,\ \log_2 m \ln\left(\frac{d}{\max\left\{1,\left(\frac{k\log_2 m}{c}\right)\right\}}\right)\right\}$$

where $c \geq 1$ is the ratio of the largest to the smallest weight among the sets in an optimal solution. This provides a smooth transition of the expected competitive ratio between "roughly" $\log_2 m \ln d$ *plus small lower order terms* for $\mathbf{WOSC}_k$ when the weights are arbitrary positive numbers to $\max\left\{5\log_2 m, \log_2 m \ln\left(\frac{d}{k\log_2 m}\right)\right\}$ for $\mathbf{OSC}_k$ when all the weights are the same.

* As a corollary of the above, for (the unweighted version) $\mathbf{OSC}_k$ for general $k$ the expected competitive ratio $\mathsf{E}\left[r(m,d,k)\right]$ decreases logarithmically with decreasing $d/k$ with a value of roughly $5\log_2 m$ in the limit[3] for all sufficiently large $k$.

- We next provide an improved analysis of $\mathsf{E}\left[r(m,d,1)\right]$ for $\mathbf{OSC}$ with better constants.

- We next provide an improved analysis of $\mathsf{E}\left[r(m,d,k)\right]$ for $\mathbf{OSC}_k$ with better constants and asymptotic limit for large $k$. The case of large $k$ is important for its application in reverse engineering of biological networks as outlined in Section 1.1. More precisely, we show that $\mathsf{E}\left[r(m,d,k)\right]$ is at most $\left(\frac{1}{2}+\log_2 m\right)\cdot\left(2\ln\frac{d}{k}+3.4\right)+1+2\log_2 m$ if $k \leq (2e)\cdot d$ and at most $1+2\log_2 m$ otherwise.

- Finally, we discuss lower bounds on competitive ratios for *deterministic algorithms* for $\mathbf{OSC}_k$ and $\mathbf{WOSC}_k$ for general $k$ using the approaches in [2]. The lower bounds obtained are $\Omega\left(\frac{\log\frac{|\mathcal{S}|}{k}\log\frac{|V|}{k}}{\log\log\frac{|\mathcal{S}|}{k}+\log\log\frac{|V|}{k}}\right)$ for $\mathbf{OSC}_k$ and $\Omega\left(\frac{\log|\mathcal{S}|\log|V|}{\log\log|\mathcal{S}|+\log\log|V|}\right)$ for $\mathbf{WOSC}_k$ for many values of the parameters.

## 1.4 Comparison With Previous Work

The structure of our algorithm is similar to and the analysis method of our algorithm is motivated by the implicit randomized algorithm (which was subsequently derandomized) in the paper *The online set cover problem* by Alon *et al.* [2].

For every set we maintain a number that will guide the process of selection; we use $\alpha p[S]$, Alon *et al.* use $w_S$. When a new element is received, and it is not covered (or sufficiently covered) yet, in both papers this number is multiplied by a constant — if the new element belongs to $S$ (in the weighted case, this number is incremented by a constant divided by $c_S$). The process of set selection is a bit different: we simply select set $S$ with probability that equals the increment of $\alpha p[S]$, while Alon *et al.* the procedure is achieving a similar effect rather indirectly — it very much looks like a de-randomization of our approach (we knew their approach when we worked on ours, so ours was a de-de-randomization).

The analysis of Alon *et al.* uses an ingenious potential function, while we use three classes of accounts. In either case, this is a form of amortized analysis. The two approaches offer distinct advantages. Alon *et al.* had a much shorter proof and could obtain a de-randomized version. As our choices were more explicitly related to Poisson trial, we applied our own versions of Chernoff bound to tighten the analysis considerably.

---

[3]Notice the similarity of this dependence of the expected competitive ratio on $d/k$ to that in our results in [7] for the offline version of the problem where we provided an approximation algorithm with an expected performance ratio of about $\max\{(1+o(1))\ln\left(\frac{d}{k}\right), 1+2\sqrt{d/k}\}$.

A fractional solution to the set cover problem is implicit in these solutions, as the "guiding numbers" can be interpreted as fractional choices, and making the "guided choices" can be interpreted as rounding. However, neither our analysis, nor that of Alon *et al.* use that fact explicitly.

## 2 A Generic Randomized Winnowing Algorithm

We first describe a generic randomized winnowing algorithm **A-Universal** below in Fig. 1. The winnowing algorithm has two scaling factors: a multiplicative scaling factor $\frac{\mu}{c_S}$ that depends on the particular set $S$ containing $i$ and another additive scaling factor $|\mathcal{S}_i|^{-1}$ that depends on the number of sets that contain $i$. These scaling factors quantify the appropriate level of "promotion" in the winnowing approach. In the next few sections, we will analyze the above algorithm for the various online set-multicover problems. The following notations will be used uniformly throughout the analysis:

- $\mathcal{J} \subseteq V$ be the set of elements received in a run of the algorithm.
- $\mathcal{T}^*$ be an optimum solution.

### 2.1 A Guided Tour — Rough Sketch of the Analysis of A-Universal for the Unweighted Case

We first sketch the overall analysis of **A-Universal** for the case when every set has cost $1$ to provide the reader an intuition behind the overall analysis of the algorithm. *Bear in mind that this analysis is neither the most precise nor the simplest, but it can be extended to the general case.* In particular we may overestimate or underestimate the constants *slightly* in the description to omit tedious details in favor of providing a better intuition.

Since the function $\mathtt{Stat}$ always returns $1$, we can remove line A4 and simplify line A6 to $p[S] \leftarrow \min(\alpha p[S] + |\mathcal{S}_i|^{-1}, 1)$.

The cost of handling an element $i$ by **A-Universal** is the number of sets that are selected. The analysis is conditional on quantity $s = \xi(i)$, where $\xi(i)$ is the sum of $\alpha p[S]$'s over $S \in \mathcal{S}_i - \mathcal{T}^*$ at the time when $i$ is received, and we take the worst case over all possible values $s$. We define event $E(b)$ that exactly $b$ sets from $\mathcal{S}_i - \mathcal{T}^*$ were already selected before element $i$ was received. Note that these selections were successes in Poisson trials that have sum of probabilities $s$, so the probability of $E(b)$ can be expressed as some $p(s, b)$, *e.g.* using Lemma 13.

The cost is split into three components: (i) selections of sets from $\mathcal{T}^*$, (ii) selections from $\mathcal{S}_i - \mathcal{T}^*$ made in lines A8-9, and (iii) selection from $\mathcal{S}_i - \mathcal{T}^*$ made in lines A11-12.

Selections of type (i) are charged to $account(\mathcal{T}^*)$, obviously the final value of this account contributes at most $1$ to the competitive ratio.

Rather than paying for the *actual* cost of selections of type (ii) and (iii), we pay the *expected* cost of these selections, and on average we will be paying enough. We estimate this cost as $s + \mathtt{deficit}$, and we pay it as follows: we charge a fixed amount $1+\psi$ to every $account(S)$ such that $S \in \mathcal{S}_i \cap \mathcal{T}^* - \mathcal{T}$, and the left-over cost is charged to $account(i)$.

The contribution of $account(S)$ to the competitive ratio is the ratio of the expected final value of $account(S)$ to the portion of $c(\mathcal{T}^*)$ attributed to $S$, and the latter happens to be $1$ (in the unweighted case!). Thus this contribution is $(1 + \psi)\beta$ where $\beta$ is the expected number of times we can charge $account(S)$. We introduce function $\Lambda(S)$ to estimate $\beta$. The initial value of $\Lambda(S) = \log_2 1 = 0$. When we charge $account(S)$ after receiving element $i$, the value of $\xi(S)$ increases from some $x$ to at least $x+x+m^{-1}$, so $mx+1$ increases to at least $2mx+2$, so $\Lambda(S)$ increases by at least $1$ — except

```
F1   function Stat(B, j)
F2       A ← ∅
F3       while (|A| < j) do        // select j least cost sets from B //
F4           S ← least cost set from B − A; insert S to A
F5       return c_S                // return the cost of the last selected set //

// definition //
D1   for (i ∈ V)
D2       S_i ← {s ∈ S : i ∈ S}

// initialization //
I1   T ← ∅                         // T is our collection of selected sets //
I2   for (S ∈ S)
I3       αp[S] ← 0                 // accumulated probability of each set //

// after receiving an element i //
A1   deficit ← k − |S_i ∩ T|       // k is the coverage factor //
A2   if deficit ≤ 0                // we need deficit more sets for i //
A3       finish the processing of i
A4   μ ← Stat(S_i − T, deficit)
A5   for (S ∈ S_i − T)
A6       p[S] ← (μ/c_S) (αp[S] + |S_i|^{-1})   // probability for this step //
A7       αp[S] ← αp[S] + p[S]      // accumulated probability //
A8       with probability min{p[S], 1}
A9           insert S to T         // randomized selection //
A10  deficit ← k − |S_i ∩ T|
A11  repeat deficit times          // greedy selection //
A12      insert a least cost set from S_i − T to T
```

Figure 1: Algorithm **A-Universal**

when $\xi[S]$ increases to $x+1$ and $S$ is deterministically selected. smaller. The average final value of $\Lambda(S)$ is at most $\log_2 m$ (cf. Lemma 4). Thus $account(S)$'s contribute roughly $(1+\psi)\log_2 m$ to the competitive ratio.

Note that there must be at least deficit many sets in $S_i \cap T^* - T$, so 1 term in $1+\psi$ surely covers the cost of selections of type (iii). If there are $b$ such sets and $s > b\psi$, we charge $s - b\psi$ to $account(i)$. To find the contribution of $account(i)$ to the competitive ratio we must ascribe part of $c(T^*)$ to $i$ and to estimate the final value of $account(i)$. If we have received $b$ elements so far, $c(T^*) \geq kb/d$, so we can ascribe $k/d$ to $i$.

Note that we make only one charge to $account(i)$. How can we estimate this charge under condition $E(j-1)$? First, because $j-1$ "incorrect" sets were already selected, deficit would be 0 if only $j-1$ "correct" sets remained unselected, so the charges are 0 *unless* we have at least $j$ unselected "correct" sets. Thus under condition $E(j-1)$, if we make any charges at all, at least $j\psi$ was charged to $account(S)$'s to cover the average cost of selections of type (ii). Thus under condition $E(b)$ we charge at most $s - (b+1)\psi$ to $account(i)$. As we estimate the probability of $E(b)$ with $p(s,b)$, we can estimate the average final value of $account(i)$ as $\sum_{j=1}^{\lfloor s/\psi \rfloor} p(s, j-1)[s-j\psi]$.

Using Lemma 13, one can show that $\psi = \max\{2, \ln(k/d)\}$ assures that $account(i)$ do not contribute more than a $\log_2 m$ factor to the competitive ratio.

## 3 An Uniform Analysis of Algorithm A-Universal

In this section, we present a uniform analysis of Algorithm **A-Universal** that applies to all versions of the online set multicover problems, *i.e.*, **OSC**, **OSC**$_k$, **WOSC** and **WOSC**$_k$. Abusing notations slightly, define $c(\mathcal{S}') = \sum_{S \in \mathcal{S}'} c_S$ for any subcollection of sets $\mathcal{S}' \subseteq \mathcal{S}$. Our bound on the competitive ratio will be influenced by the parameter $\kappa$ defined as: $\kappa = \min_{i \in \mathcal{J} \ \& \ S \in \mathcal{S}_i \cap \mathcal{T}^*} \left\{ \frac{c(\mathcal{S}_i \cap \mathcal{T}^*)}{c_S} \right\}$. It is easy to check that $\kappa = \begin{cases} 1 & \text{for } \mathbf{OSC} \\ k & \text{for } \mathbf{OSC}_k \\ \geq 1 & \text{for } \mathbf{WOSC} \text{ and } \mathbf{WOSC}_k \end{cases}$. The main result proved in this section is the following theorem.

**Theorem 1** *The expected competitive ratio* $\mathsf{E}[r(m, d, k)]$ *of Algorithm* **A-Universal** *is at most*

$$1 + \log_2 m \times \max\left\{5, \ 2 + \ln \frac{d}{\kappa \log_2 m}\right\}$$

**Corollary 2**
(a) *For* **OSC**, **WOSC** *and* **WOSC**$_k$*, setting* $\kappa = 1$ *we obtain* $\mathsf{E}[r(m, d, k)]$ *to be at most* $\log_2 m \ln d$ *plus lower order terms.*
(b) *For* **OSC**$_k$*, setting* $\kappa = k$*, we obtain* $\mathsf{E}[r(m, d, k)]$ *to be at most*

$$1 + \log_2 m \times \max\left\{5, \ \left(2 + \ln \frac{d}{k \log_2 m}\right)\right\}$$
$$\approx \log_2 m \times \max\left\{5, \ \ln \frac{d}{k \log_2 m}\right\}$$

(c) *Let* $c \geq 1$ *is the ratio of the largest to the smallest weight among the sets in an optimal solution. Then, setting* $\kappa = \max\left\{1, \frac{k}{c}\right\}$*, we obtain* $\mathsf{E}[r(m, d, k)]$ *to be at most*

$$1 + \log_2 m \times \max\left\{5, \ \left(2 + \ln \frac{d}{\max\{1, \frac{k}{c}\} \log_2 m}\right)\right\}$$
$$\approx \log_2 m \times \max\left\{5, \ \ln\left(\frac{d}{\max\{1, \left(\frac{k \log_2 m}{c}\right)\}}\right)\right\}$$

In the next few subsections we prove the above theorem.

### 3.1 The overall scheme

We first roughly describe the overall scheme of our analysis. The average cost of a run of **A-Universal** is the sum of average costs that are incurred when elements $i \in \mathcal{J}$ are received. We will account for these costs by dividing these costs into three parts $\text{cost}_1 + \sum_{i \in \mathcal{J}} \text{cost}_2^i + \sum_{i \in \mathcal{J}} \text{cost}_3^i$ where:

**cost**$_1 \leq c(\mathcal{T}^*)$ upper bounds the *total* cost incurred by the algorithm for selecting sets in $\mathcal{T} \cap \mathcal{T}^*$.

**cost**$_2^i$ is the cost of selecting sets from $\mathcal{S}_i - \mathcal{T}^*$ in line A9 for each $i \in \mathcal{J}$.

**cost**$_3^i$ is the cost of selecting sets from $\mathcal{S}_i - \mathcal{T}^*$ in line A12 for each $i \in \mathcal{J}$.

We will use the accounting scheme to count these costs by creating the following three types of accounts:

$account(\mathcal{T}^*)$;
$account(S)$ for each set $S \in \mathcal{T}^* - \mathcal{T}$;
$account(i)$ for each received element $i \in \mathcal{J}$.

$cost_1$ obviously adds at most 1 to the average competitive ratio; we will charge this cost to $account(\mathcal{T}^*)$. The other two kinds of costs, namely $cost_2^i + cost_3^i$ for each $i$, will be distributed to the remaining two accounts. Let $D = \frac{d}{\kappa \log_2 m}$. The distribution of charges to these two accounts will satisfy the following:

- $\sum_{i \in \mathcal{J}} account(i) \leq \log_2 m \cdot c(\mathcal{T}^*)$. This claim in turn will be satisfied by:
  - dividing the optimal cost $c(\mathcal{T}^*)$ into pieces $c_i(\mathcal{T}^*)$ for each $i \in \mathcal{J}$ such that $\sum_{i \in \mathcal{J}} c_i(\mathcal{T}^*) \leq c(\mathcal{T}^*)$; and
  - showing that, for each $i \in \mathcal{J}$, $account(i) \leq \log_2 m \cdot c_i(\mathcal{T}^*)$.

- $\sum_{S \in \mathcal{T}^*} account(S) \leq \log_2 m \cdot \max\{4, \ln D + 1\} \cdot c(\mathcal{T}^*)$.

This will obviously prove an expected competitive ratio of at most the maximum of $1 + 5(1 + \log_2 m)$ and $1 + (1 + \log_2 m)(2 + \ln D)$, as promised.

We will perform our analysis from the point of view of each received element $i \in \mathcal{J}$. To define and analyze the charges we will define several quantities:

| | |
|---|---|
| $\mu(i)$ | the value of $\mu$ calculated in line A4 after receiving $i$ |
| $\xi(i)$ | the sum of $\alpha p[S]$'s over $S \in \mathcal{S}_i - \mathcal{T}^*$ at the time when $i$ is received |
| $a(i)$ | $\|\mathcal{T} \cap \mathcal{S}_i - \mathcal{T}^*\|$ at the time when $i$ is received |
| $\Lambda(S)$ | $\log_2(m \cdot \alpha p[S] + 1)$ for each $S \in \mathcal{S}$; it changes during the execution of **A-Universal** |

Finally, let $\Delta(X)$ denote the amount of change (increase or decrease) of a quantity $X$ when an element $i$ is processed.

## 3.2 The role of $\Lambda(S)$

We will ensure the *invariant $account(S) \leq \max\{4, \ln D + 1\} \cdot \Lambda(S) \cdot c_S$* for every $S \in \mathcal{T}^*$. We will simply not accept larger charges to the accounts of sets than this invariant allows. This invariant is useful because we will prove a universal upper bound $U$ on the expected final value of $\Lambda(S)$, and thus the contribution of the accounts of sets to the expected competitive ratio will be $\max\{4, \ln D + 1\} \cdot U$.

**Definition 3** *When we determine the charges to accounts made when element $i$ is received, we classify sets from $\mathcal{S}_i \cap \mathcal{T}^* - \mathcal{T}$ as heavy if $c_S \geq \mu(i)$ and light otherwise.*

When $i$ is received we charge accounts of $S \in \mathcal{T}^* \cap \mathcal{S}_i - \mathcal{T}$ in the following manner:

- for a light set, $\Delta(account(S)) = c_S$ while we can show that $\Delta(\Lambda(S)) > 1$ and
- for a heavy set $\Delta(account(S)) = \max\{4, \ln D + 1\}\mu(i)$ while $\Delta(\Lambda(S)) \geq \mu(i)/c_S$.

The above estimates of $\Delta(\Lambda(S))$ are easy to show: in lines A6-7 we increment $\alpha p[S] + m^{-1}$ with

$$\frac{\mu(i)}{c_S}(\alpha p[S] + |S_i|^{-1}) \geq \frac{\mu(i)}{c_S}(\alpha p[S] + m^{-1}),$$

which increments $\Lambda(S) = \log_2(\alpha p[S] + |S_i|^{-1}) - \log_2 m$ by at least $\log_2(1 + \mu(i)/c_S)$; for a light set this increment is at least $\log_2 2 = 1$, and for a heavy set we have $\mu(i)/c_S \leq 1$, and we use the following fact:
$$\log_2(1+x) \geq x \text{ for } x \leq 1.$$

Of course, such an approach makes sense only if we can prove an upper bound on $E[\Lambda(S)]$. Note that in step A6 we may calculate a value of $p[S]$ that is larger than 1.

We analyze $E[\Lambda(S)]$ from the following point of view: consider a fixed sequence of $p[S]$ over the execution of the algorithm; each time $p[S] > 0$ there is a chance that $S$ gets selected and this is the last step when $\Lambda(S)$ increases. Our bound will hold true for every possible sequence.

**Lemma 4** $E[\Lambda(S)] \leq \log_2 m$ for $m \geq 7$.

**Proof.** We want to find the expected final value of $\Lambda(S) = \log_2(m \cdot \alpha p[S]+1) = \log_2 m + \log_2(\alpha p[S] + m^{-1})$. It is a function of the sequence of probabilities, say $p_1, p_2, \ldots$, that $p[S]$ computed when elements of $S$ were received.

We will be working with sequences formed from possible sequences of probabilities by deleting an initial part; let the sum of this initial part and $m^{-1}$ is $z$. We define $\beta p_i = z + \sum_{j=1}^{i-1} p_j$ which stands for the value of $\alpha p[S] + m^{-1}$ in line A6 when we compute $p_i$. We say that $\vec{p} = (p_1, p_2, \ldots)$ is $z$-legal if for $i \geq 1$ we have $0 \leq p_i \leq \beta p_i$, and if $p_i \geq 1$ then $p_i$ is the last term of $\vec{p}$. Let $\text{tail}(\vec{p}) = (p_2, \ldots)$.

We define $F(z, \vec{p})$ as follows. If $\vec{p}$ is an empty sequence then $F(z, \vec{p}) = 0$, otherwise

$$F(z, \vec{p}) = p_1 \log_2(p_1 + z) + (1 - p_1) F(z + p_1, \text{tail}(\vec{p})) \quad (*)$$

In turn, $F(z)$ is the supremum value of $F(z, \vec{p})$ over all $z$-legal sequences. Our goal is to show that $F(1/m) < 0$ for $m \geq 7$.

We first show that if the supremum defining $F(z)$ is limited to infinite sequences, then it is finite. By repetitively applying formula (*) we get

$$F(z, p) = \sum_{i=1}^{\infty} \prod_{j=1}^{i-1}(1 - p_j) p_i \log_2(\beta p_i + p_i) < e^z \int_z^{\infty} e^{-x} \log_2(x+1) dx$$

where the summation can be converted to an integral as follows: $p_i$ can be a sum of $dx$'s over an interval of length $p_i$, say from $\beta p_i$ to $\beta p_{i+1}$, the product can be the probabilistic density function that can be bounded from above with $e^{z-x}$ and $\log_2$ can be the function that we compute expectation of, and it can be estimated from above with $\log_2(x+1)$; this justifies the estimate with of $F(z, \vec{p})$ with a convergent integral.

Next we show that for $z \geq \log_2 e$ we have $F(z) = F(z, (z)) = 1 + \log_2 z$. Suppose that $F(z) > 1 + \log_2 z$. Then for some finite $\vec{p}$ and for some $z \geq \log_2 e$ we have $F(z, \vec{p}) > F(z, (z)) = 1 + \log_2 z$. Consider a shortest such sequence. Because of (*) we can conclude that $\vec{p}$ has length 2, since otherwise $F(z + p_1, \text{tail}(\vec{p})) \leq F(z + p_1, (z + p_1))$, but in that case we can replace $\text{tail}(\vec{p})$ with the single term $z + p_1$. So we can assume that $\vec{p} = (x, z+x)$ for some $x > 0$. Then we have

$$F(z, \vec{p}) = x \log_2(z+x) + (1-x)\log_2(z+x+z+x) > 1 + \log_2 z$$
which implies
$$x \log_2(z+x) + (1-x)(1 + \log_2(z+x)) > 1 + \log_2 z$$
which implies
$$x \left(\log_2 z + \log_2 \tfrac{z+x}{z}\right) + (1-x)\left(1 + \log_2 z + \log_2 \tfrac{z+x}{z}\right) > 1 + \log_2 z$$
which implies
$$\log_2 \tfrac{z+x}{z} > x$$

The latter is not possible, because for $x \geq z \geq \log_2 e$ the derivative of the left-hand-side is $\frac{\log_2 e}{z+x} \leq 1$, while the derivative of the right-hand-side is 1.

In a $z$-legal sequence $\vec{p}$ we have $p_1 \leq \min\{1, z\}$. As the third observation we can show that if $\beta p$ has more than one term, then $p_1 + p_2 > \min\{1, z\}$, otherwise we increase $F(z, \vec{p})$ when we coalesce the first two terms of $\vec{p}$ into one. Let $p_1 = x, p_2 = y, p_1 + p_2 = p$, we have

$$x \log_2(z+x) + (1-x)y \log_2(z+p) + (1-x)(1-y)F(z+p) < p \log_2(z+p) + (1-p)F(z+p)$$
which implies
$$x \left(\log_2 \tfrac{z+x}{z+p} + \log_2(z+p)\right) + (1-x)y \log_2(z+p) + xy F(z+p) < p \log_2(z+p)$$
which implies
$$x \log_2 \tfrac{z+x}{z+p} - xy \log_2(z+p) + xy F(z+p) < 0$$
which implies
$$F(z+p) < \log_2(z+p) + \tfrac{1}{y} \log_2\left(1 + \tfrac{y}{z+p-y}\right)$$

Because we always have $F(z) \leq \log_2(z) + 1$, it suffices to show that $\frac{1}{y}\log_2(1 + \frac{y}{z+p-y}) > 1$. This follows from the fact that for $x < \log_2 e$ the derivative of $\log_2 x$ is larger than 1.

The methods used to show the last two fact allow to characterize the optimal (or worst case) sequences: if $z \geq \log_2 e$, use 1-term sequence consisting of $z$, otherwise start from $\min\{z, 1, \log e - z\}$.

As a consequence, if $\frac{1}{2}\log_2 e \leq z \leq \log_2 e$ then $F(z) = F(z, (\log_2 e - z, \log_2 e)) = \log_2 \log_2 e + 1 - \log_2 e + z$, and for $z \leq \frac{1}{2}\log_2 e$ we know that $F(z) = z \log_2(2z) + (1-z)F(2z)$. It is easy to see that for $F(z/2) < F(z)$, and we can compute the values of $F(1/m)$ for $m = 2, 3, \ldots, 7$:

| $m$ | 1 | 2 | 3 | 4 | 5 | 6 | 7 | 8 |
|---|---|---|---|---|---|---|---|---|
| $F(1/m)$ | 1.086 | 0.543 | 0.397 | 0.157 | 0.120 | 0.067 | $-0.016$ | $-0.112$ |

❑

Observe that it is very easy to show the competitive ratio of $m$, so for $m = 1$ it makes no sense to discuss the competitive ratio, while for $1 < m \leq 16$, since $4 \log_2 m \geq m$, the upper bound we are proving is trivial.

### 3.3 Charges due to the costs of line A12

When we make greedy selections in line A12, there are at least deficit many sets in $\mathcal{S}_i \cap \mathcal{T}^* - \mathcal{T}$; we can order them according to their costs, say $S_1, S_2, \ldots$; and let $c_{S_i} = a_i$. Because we could make greedy selections of these sets, the costs of actual selections cannot be larger, so if these costs are ordered $b_1 \leq \ldots \leq b_{\text{deficit}}$, we have $b_i \leq a_i$ for $i = 1, \ldots, \text{deficit}$.

Therefore we can charge $b_i$ to $account(S_i)$ and the expected sum of such charges made to each $account(S)$ is at most $c_S \cdot \log_2 m$. Therefore these charges contribute $\log_2 m$ to the expected competitive ratio.

## 3.4 Charges due to the costs of line A9

The expected sum of charges due to the costs of line A9 equals $\mu(i)\xi(i) + \mu(i)$: every set from $\mathcal{S}_i - \mathcal{T} - \mathcal{T}^*$ contributes, regardless of its weight, $\mu(i)(\alpha p[S] + |\mathcal{S}_i|^{-1})$, $\alpha p[S]$ terms add to $\xi(i)$, while $|\mathcal{S}_i|^{-1}$ terms add to 1. We will refer to these two terms as A9a charges and A9b charges.

A9b charges will be given to an arbitrary account of a heavy set (in the worst case, there is only one).

A9a charges are distributed among the accounts of heavy sets and *account*(i). The idea is the following: we will fix the A9a charge to each heavy set account to some $\psi$ such that the contribution of these charges to the competitive ratio will be exactly $\mu(i)\psi$. We estimate the number of the heavy sets as follows.

**Lemma 5** *There are at least $a(i) + 1$ heavy sets.*

**Proof.** Our assumption is that at the time $i$ is received, $a(i)$ sets from $\mathcal{S}_i - \mathcal{T}^*$ are already selected to $\mathcal{T}$. Thus when we compute $\mu(i)$ in a call to $\text{Stat}(\mathcal{S}_i - \mathcal{T})$ in line A4 we can form set $\mathcal{A}$ from $\mathcal{S}_i \cap \mathcal{T}^*$ after excluding $a(i)$ sets with the largest cost. Would we do that, $\mu(i)$ would become the largest cost in $\mathcal{S}_i \cap \mathcal{T}^* - \mathcal{T}$, after excluding $a(i)$ costs that are yet larger, so we indeed have at least $a(i)$ sets of cost $\mu(i)$ or more — hence heavy. When we include other sets in $\mathcal{A}$ as well, the value of $\mu(i)$ can only decrease, and then the number of heavy sets can only increase. ❑

Therefore at most $\mu(i)(\xi(i) - (a(i) + 1)\psi)$ will be charged to *account*(i). Thus we need to show that $E[\xi(i) - (a(i) + 1)\psi]$ is sufficiently small.

The intuition is that when $\xi(i)$ is small, the charges cannot be made, and when $\xi(i)$ is large, the average value of $a(i)$ is equally large and thus the probability of making charges is sufficiently small to assure a very small average value.

In the next subsection we analyze these probabilities, but it is easy to see that the higher $\psi$, the smaller $E[\xi(i) - (a(i) + 1)\psi]$. We want to set the average charge to *account*(i) in such a way that the expected contribution of these accounts to the competitive ratio is at most $\log_2 m$. So the question is: how large portion of $c(\mathcal{T}^*)$ can we attribute to element $i$?

To simplify our calculations, we rescale the costs of sets so $\mu(i) = 1$ and thus $c_S \geq 1$ for each heavy set $S$ and the sum of charges due to line A9 is simply $\xi(i)$.

We associate with $i$ a piece $c_i(\mathcal{T}^*)$ of the optimum cost $c(\mathcal{T}^*)$:

$$c_i(\mathcal{T}^*) = \sum_{S \in \mathcal{S}_i \cap \mathcal{T}^*} c_S/|S| \geq \frac{1}{d}c(\mathcal{S}_i \cap \mathcal{T}^*) \geq \frac{\kappa}{d}\mu(i) = \kappa/d.$$

It is then easy to verify that

$$\sum_{i \in \mathcal{J}} c_i(\mathcal{T}^*) \leq \sum_{i \in \mathcal{J}} \frac{1}{d}c(\mathcal{S}_i \cap \mathcal{T}^*) \leq c(\mathcal{T} \cap \mathcal{T}^*) \leq c(\mathcal{T}^*)$$

Thus we can charge *account*(i) in such a way that on average it receives $(\kappa/d)\log_2 m$, and let $D^{-1} = (\kappa/d)\log_2 m$. In the next subsection, we find a sufficiently high value of $\psi$ to make it so. For now observe that the competitive ratio will be $1 + (3 + \psi)\log_2 m$: 1 for the charges to *account*($\mathcal{T}^*$), $\log_2 m$ for the charges due to line A12, $\log_2 m$ for the charges to *account*(i)'s, $\log_2 m$ for A9b charges and $\psi\log_2 m$ for A9a charges.

## 3.5 Split of A9a charges between $i$ and the heavy sets

In this section we prove that for $\psi = \max\{2, \ln D - 1\}$ we have $E[\xi(i) - (a(i) + 1)\psi] \leq D^{-1}$.

Define
$$\mathcal{E}(i, b) = \begin{cases} 1 & \text{if } a(i) \leq b \\ 0 & \text{otherwise} \end{cases}$$

Let $\mathrm{charge}(i, \psi, \ell, x)$ be the formula for the charge to *account*($i$) assuming we use $\psi$ with $\ell\psi \leq x = \xi(i) \leq (\ell+1)\psi$. We can estimate $\mathrm{charge}(i, \psi, \ell, x)$ in the following manner:

- If $\mathcal{E}(i, \ell - 1) = 1$, then $a(i) + 1 = \ell$, the total charge to all the heavy sets is $\ell\psi$ and thus we have to charge *account*($i$) with $x - \ell\psi$.

- if $\mathcal{E}(i, \ell - 2) = 1$ then we also have $\mathcal{E}(i, \ell - 1) = 1$, so we charged *account*($i$) with $x - \ell\psi$ already, but we need to charge *account*($i$) with an additional amount of $\psi$.

- Continuing in a similar manner, it follows that for each $b \leq \ell - 2$, if $\mathcal{E}(i, b) = 1$ we charge *account*($i$) with an additional amount of $\psi$.

Thus we get the following estimate:

$$E[\mathrm{charge}(i, \psi, \ell, x)] = \Pr[\mathcal{E}(i, \ell - 1) = 1] \cdot (x - \ell\psi) + \psi \sum_{j=0}^{\ell-2} \Pr[\mathcal{E}(i, j) = 1].$$

Since $\psi(a(i) + 1) < \xi(i)$ and $\psi \geq 2$, $a(i) + 1$ is less than $\frac{1}{2}\xi(i)$. Thus, we can use Lemma 13 with $X = x = \xi(i)$ and $a = j$ to obtain $\Pr[\mathcal{E}(i, j) = 1] < e^{-x}\frac{x^j}{j!}$ for $j = \ell - 1, \ell - 2, \ldots, 0$. Let $C(\psi, \ell, x)$ be the estimate of of $E[\mathrm{charge}(i, \psi, \ell, x)]$ thus obtained:

$$C(\psi, \ell, x) = e^{-x}\left(\frac{x^{\ell-1}}{(\ell-1)!}(x - \ell\psi) + \psi \sum_{j=0}^{\ell-2} \frac{x^j}{j!}\right).$$

**Lemma 6** *If $\psi \geq 2$, $x \geq 1$ and $\ell = \lfloor x/\psi \rfloor \geq 1$ then $C(\psi, \ell, x) \leq e^{-(\psi+1)}$.*

**Proof.** We first consider the case of $\ell = 1$. Because $\mathcal{E}(i, -1)$ is not possible, $\mathrm{charge}(i, \psi, 1, x) = \mathcal{E}(i, 0)(x - \psi)$ and $C(\psi, 1, x) = e^{-x}(x - \psi)$. Now since $\frac{\partial}{\partial x}C(\psi, 1, x) = e^{-x}(-x + \psi + 1)$, $C(\psi, 1, x)$ is maximized for $x = \psi + 1$ with a maximum value of $e^{-(\psi+1)}$.

For $\ell \geq 2$ the summation part of the formula for $C(\psi, \ell, x)$ is non-trivial; in that case one can calculate that
$$\frac{\partial}{\partial x}C(\psi, \ell, x) = e^{-x}\frac{x^{\ell-2}}{(\ell-1)!}(-x^2 + \ell(\psi + 1)x - (\ell^2 - 1)\psi).$$

As we see, this derivative is a product of a positive function with a trinomial. This trinomial has the maximum for $x = \ell(\psi+1)/2$, so in our range, $\ell\psi \leq x \leq (\ell+1)\psi$, it is decreasing. For $x = \ell\psi$ the value of the trinomial is $\psi > 0$, and for $x = \ell\psi + 2/\ell$ the value of the trinomial is $2 - \psi - 4\ell^{-2} < 0$. Therefore the maximum must occur in the interval between $\ell\psi$ and $\ell\psi + 2/\ell$ and it will suffice to prove our claim in this range.

For $x = \ell\psi + z$ with $0 < z < 2/\ell$ the inequality we want to prove is equivalent to

$$\mathrm{LHS} = \frac{(\ell\psi + z)^{\ell-1}}{(\ell-1)!}z + \psi \sum_{j=0}^{\ell-2} \frac{(\ell\psi + z)^j}{j!} \leq e^{(\ell-1)\psi - 1 + z} = \mathrm{RHS} \qquad (1)$$

Suppose that (1) is true for some $\psi$; then for $\psi' = \psi + \varepsilon$ RHS increases by a factor of $e^{(\ell-1)\varepsilon}$, while each monomial $\frac{(\ell\psi+z)^j}{j!}$, for $j = 0, 1, \ldots, \ell-1$, increases by a factor of $\left(1 + \frac{\varepsilon}{\psi+\frac{z}{\ell}}\right)^j \leq \left(1 + \frac{\varepsilon}{\psi}\right)^{\ell-1} < e^{(\ell-1)\frac{\varepsilon}{\psi}}$ and thus the entire LHS increases by a factor of at most $\psi e^{(\ell-1)\frac{\varepsilon}{\psi}} < e^{(\ell-1)\varepsilon}$. Because LHS increases less that RHS, the inequality for $\psi$ implies that for $\psi + \varepsilon$ and thus for every higher value. For this reason it suffices to prove the inequality for $\psi = 2$ and for $\ell\psi < x < \ell\psi + 2/\ell$ (thus, for $0 < z < 2/\ell$). For $\psi = 2$, our claim is reduces to

$$\text{LHS} = \frac{(2\ell+z)^{\ell-1}}{(\ell-1)!}z + 2\sum_{j=0}^{\ell-2}\frac{(2\ell+z)^j}{j!} \leq e^{2\ell-3+z} = \text{RHS}$$

For convenience, let $y = 2\ell + z$. Thus, we need to prove

$$\text{LHS} = \frac{y^{\ell-1}}{(\ell-1)!}(y - 2\ell) + 2\sum_{j=0}^{\ell-2}\frac{y^j}{j!} \leq e^{y-3} = \text{RHS}$$

subject to $2\ell < y < 2\ell + \frac{2}{\ell}$. Since $\ell \geq 2$, $y < 2\ell + \frac{2}{\ell} < 2(\ell+1)$ and thus $y - 2\ell < 2$. Thus LHS $< 2\sum_{j=0}^{\ell-1}\frac{y^j}{j!}$, and since, by the well-known series expansion, $e^y = \sum_{j=0}^{\infty}\frac{y^j}{j!}$ it suffices to show that

$$2e^3 \sum_{j=0}^{\ell-1} T_j \leq \sum_{j=0}^{\infty} T_j$$

for $\ell \geq 2$, $2\ell < y < 2\ell + \frac{2}{\ell}$ and $T_j = \frac{y^j}{j!}$. First, we verify by induction that $T_j \geq \sum_{i=0}^{j-1} T_i$ for $1 \leq j \leq \ell$. Note that for $1 \leq j \leq \ell$, $T_j/T_{j-1} = y/j > 2$. For the basis case of $j = 1$, it is therefore obvious. Otherwise, $T_j > 2T_{j-1} > T_{j-1} + \sum_{i=0}^{j-2} T_i = \sum_{i=0}^{j-1} T_i$ by inductive hypothesis. Thus, it suffices to show that

$$2e^3 T_\ell \leq \sum_{j=0}^{\infty} T_j$$

For $\ell + 1 \leq j \leq 2\ell$, $T_j/T_{j-1} = y/j > 1$. Thus, $\sum_{j=0}^{\infty} T_j \geq \ell T_\ell$, and thus it suffices to show that $2e^3 T_\ell \leq \ell \cdot T_\ell$ which holds provided $\ell \geq 2e^3 \approx 40.17$. Thus, the claim holds for $\ell > 40$.

For $2 \leq \ell \leq 40$ and $\psi = 2$, we can verify our claim by easy numerical calculation. Notice that we just need to verify $C(2, \ell, x_0) \leq e^{-3}$ where $x_0$ is the real root of the quadratic function $f(x) = -x^2 + 3\ell x - 2(\ell^2 - 1)$ that lies in the range $2\ell < x < 2\ell + 2/\ell$. By numerical calculation, one can tabulate the results as shown in Table 1 and verify that $C(2, \ell, x_0) < 0.049 < e^{-3}$. ❑

Now, since $\psi = \max\{2, \ln D - 1\} \geq 2$ we conclude using Lemma 6 that the average charge to $account(i)$ is at most $e^{-\ln D} = D^{-1}$.

## 4 Improved Analysis of Algorithm A-Universal for Unweighted Cases

In this section, we provide improved analysis of the expected competitive ratios of Algorithm **A-Universal** or its minor variation for the unweighted cases of the online set multicover problems. These improvements pertain to providing improved constants in the bound for $E[r(m, d, k)]$. The following notations will be used in this section:

| $\ell$ | $x_0$ | $C(2, \ell, x_0)$ |
|---|---|---|
| 40 | 80.049938 | 0.000000267802482750 |
| 39 | 78.051215 | 0.000000367770130466 |
| 38 | 76.052559 | 0.000000505162841918 |
| 37 | 74.053975 | 0.000000694037963620 |
| 36 | 72.055470 | 0.000000953753092710 |
| 35 | 70.057050 | 0.000001310973313578 |
| 34 | 68.058722 | 0.000001802442476141 |
| 33 | 66.060495 | 0.000002478811076980 |
| 32 | 64.062378 | 0.000003409926108503 |
| 31 | 62.064382 | 0.000004692144890365 |
| 30 | 60.066519 | 0.000006458452590756 |
| 29 | 58.068802 | 0.000008892465898008 |
| 28 | 56.071247 | 0.000012247826675415 |
| 27 | 54.073872 | 0.000016875076361489 |
| 26 | 52.076697 | 0.000023258920058581 |
| 25 | 50.079746 | 0.000032069930688629 |
| 24 | 48.083046 | 0.000044236337186173 |
| 23 | 46.086630 | 0.000061043767052413 |
| 22 | 44.090537 | 0.000084273925651732 |
| 21 | 42.094810 | 0.000116397546202183 |
| 20 | 40.099505 | 0.000160843029165595 |
| 19 | 38.104686 | 0.000222370693445282 |
| 18 | 36.110434 | 0.000307594429791974 |
| 17 | 34.116844 | 0.000425709065373619 |
| 16 | 32.124038 | 0.000589504628397967 |
| 15 | 30.132169 | 0.000816780125566277 |
| 14 | 28.141428 | 0.001132311971151022 |
| 13 | 26.152067 | 0.001570588251431389 |
| 12 | 24.164414 | 0.002179590204991318 |
| 11 | 22.178908 | 0.003025980931596380 |
| 10 | 20.196152 | 0.004202124182703906 |
| 9 | 18.216991 | 0.005835328094363729 |
| 8 | 16.242641 | 0.008099376451161879 |
| 7 | 14.274917 | 0.011227174827357965 |
| 6 | 12.316625 | 0.015519482245119539 |
| 5 | 10.372281 | 0.021333034990024608 |
| 4 | 8.449490 | 0.028995023101223379 |
| 3 | 6.561553 | 0.038468799615120751 |
| 2 | 4.732051 | 0.048129928161242959 |

Table 1: Verification of $C(2, \ell, x_0) < e^{-3}$ for $2 \leq \ell \leq 40$.

$\sigma p[i] = \sum_{S \in \mathcal{S}_i} p[S];$
$\sigma \alpha p[i] = \sum_{S \in \mathcal{S}_i} \alpha p[S];$
$\widetilde{T}$ is the set of elements of $T$ for which line A5 was executed.

## 4.1 Improved performance bounds for OSC

**Theorem 7** $E[r(m, d, 1)] \leq \begin{cases} \log_2 m \ln d, & \text{if } m > 15 \\ \left(\frac{1}{2} + \log_2 m\right)(1 + \ln d), & \text{otherwise} \end{cases}$

In the rest of the section, we prove the above theorem via a series of claims. Note that for **OSC** we substitute $\mu = c_S = k = 1$ in the psuedocode of Algorithm **A-Universal** and that $\text{deficit} \in \{0, 1\}$.

**Lemma 8** *For any* $T \in \mathcal{T}^*$, $E\left[|\widetilde{T}|\right] \leq \begin{cases} \frac{1}{2} + \log_2 m, & \text{if } m \leq 7 \\ \log_2 m, & \text{otherwise} \end{cases}$

**Proof.** We can use the proof of Lemma 4 with small exceptions. The sequence of probabilities that are computed are always doubling the previous one, so for $z \geq 1$ we always use probability 1 and as the result, $F(z) = \log_2 z + 1$, and thus $F(1) = 1$. Similarly, for $\frac{1}{2} \leq z \leq 1$ we have $F(z) = z(\log_2 z + 1) + (1-z)(\log_2 z + 2) = \log_2 z + 2 - z$, and thus $F(z) = \frac{1}{2}$. In turn, $E\left[|\widetilde{T}|\right] = \log_2 m + F(1/m)$, so for $m \geq 2$ we have $E\left[|\widetilde{T}|\right] \leq \log_2 m + \frac{1}{2}$ and for $m \geq 7$ we have $E\left[|\widetilde{T}|\right] \leq \log_2 m$. ❑

Obviously $E[|\mathcal{T}|]$ is equal to the sum of probabilities used in line A12 plus the number of times we execute line A12. Let $\xi(i)$ be the value of $\sigma \alpha p[i]$ *at the time* the algorithm receives element $i$ as the input. If the test of line A2 is false, the sum of probabilities used in line A6 is $\xi(i) + 1$, while by Lemma 13 with $\alpha = 0$ line A12 is executed with probability at most $\frac{1}{e} < 0.37$, so the contribution of $i$ to the expected cost is smaller than $\xi(i) + 1.37$.

**Lemma 9** *For* $T \in \mathcal{T}^*$, *if* $|\widetilde{T}| > 0$ *then* $E\left[\sum_{i \in \widetilde{T}} \xi(i)\right] < E\left[|\widetilde{T}|\right]\left(\ln |T| - \ln E\left[|\widetilde{T}|\right]\right)$.

**Proof.** Before the condition in line A2 is evaluated for element $i$ the algorithm performs independent random selections of sets from $\mathcal{S}_i$ with the sum of probabilities of success equal to $\xi(i)$. By Lemma 13 with $\alpha = 0$ the probability that all these selections fail, and thus the test in line A2 is false, is $\Pr\left[i \in \widetilde{T}\right] < e^{-\xi(i)}$. Let $\Gamma$ be a parameter to be established later, and let $\zeta(i) = \max\{0, \xi(i) - \ln|T| + \Gamma\}$. Clearly,

$$E\left[\sum_{i \in \widetilde{T}} \xi(i)\right] \leq E\left[|\widetilde{T}|\right](\ln|T| - \Gamma) + \sum_{i \in T} \Pr\left[i \in \widetilde{T}\right]\zeta(i)$$

Let $T' = \{i \in T : \zeta(i) > 0\}$. Then

$$\sum_{i \in T} \Pr\left[i \in \widetilde{T}\right]\zeta(i) \leq \sum_{i \in T'} e^{-\zeta(i) - \ln|T| + \Gamma}\zeta(i) = |T|^{-1} e^{\Gamma} \sum_{i \in T'} e^{-\zeta(i)}\zeta(i) < e^{\Gamma - 1}.$$

where the last inequality follows from Fact 2 and $T' \subseteq T$. Thus,

$$E\left[\sum_{i \in \widetilde{T}} \xi(i)\right] \leq E\left[|\widetilde{T}|\right]\left(\ln|T| - \Gamma + \frac{e^{\Gamma - 1}}{E\left[|\widetilde{T}|\right]}\right)$$

We can use $\Gamma = 1 + \ln \mathsf{E}\left[|\widetilde{\mathsf{T}}|\right]$ to get the desired estimate. ❑

Now, we are ready to finish the proof of the claim on $\mathsf{E}\left[r(m, d, 1)\right]$ in the theorem.

$$\begin{aligned}
\mathsf{E}\left[r(m, d, 1)\right] = \frac{\mathsf{E}[|\mathcal{T}|]}{|\mathcal{T}^*|} &< \frac{\sum_{\mathsf{T} \in \mathcal{T}^*} \mathsf{E}[\sum_{i \in \widetilde{\mathsf{T}}} \xi(i) + 1.37]}{|\mathcal{T}^*|} \\
&< \frac{\sum_{\mathsf{T} \in \mathcal{T}^*} \mathsf{E}[|\widetilde{\mathsf{T}}|]\left(\ln|\mathsf{T}| - \ln \mathsf{E}[|\widetilde{\mathsf{T}}|] + 1.37\right)}{|\mathcal{T}^*|} \quad \text{(by Lemma 9)} \\
&= \mathsf{E}\left[|\widetilde{\mathsf{T}}|\right]\left(\ln|\mathsf{T}| - \ln \mathsf{E}\left[|\widetilde{\mathsf{T}}|\right] + 1.37\right)
\end{aligned}$$

The last quantity is an increasing function of $\mathsf{E}\left[|\widetilde{\mathsf{T}}|\right]$, so we can replace it with its overestimate. For every $m \geq 2$ we can use estimate $\mathsf{E}\left[|\widetilde{\mathsf{T}}|\right] \leq 0.5 + \log_2 m$ and the fact that $\ln(0.5 + \log_2 2) > 0.37$. For $m \geq 16$ we can use estimate $\mathsf{E}\left[|\widetilde{\mathsf{T}}|\right] \leq \log_2 m$ and the fact that $\ln \log_2 16 > 1.37$.

## 4.2 Improved performance bounds for $\mathbf{OSC}_k$

Note that for $\mathbf{OSC}_k$ we substitute $\mu = c_S = 1$ in the psuedocode of Algorithm **A-Universal** and that $\mathtt{deficit} \in \{0, 1, 2, \ldots, k\}$. For improved analysis, we change Algorithm **A-Universal** *slightly*, namely, line A6 (with $\mu = c_S = 1$)

    A6    $p[S] \leftarrow \min\left\{\left(\alpha p[S] + |\mathcal{S}_i|^{-1}\right), 1\right\}$ // probability for this step //

is changed to

    A6'    $p[S] \leftarrow \min\left\{\left(\alpha p[S] + \mathtt{deficit} \cdot |\mathcal{S}_i|^{-1}\right), 1\right\}$ // probability for this step //

**Theorem 10** *With the above modification of Algorithm* **A-Universal**,
$$\mathsf{E}\left[r(m, d, k)\right] \leq \begin{cases} \left(\frac{1}{2} + \log_2 m\right) \cdot \left(2\ln\frac{d}{k} + 3.4\right) + 1 + 2\log_2 m & \textit{if } k \leq (2e) \cdot d \\ 1 + 2\log_2 m & \textit{otherwise} \end{cases}$$

We now proceed with the proof of the above theorem. As before, $\mathcal{T}^*$ is an optimal solution and for $\mathsf{T} \in \mathcal{T}^*$ we define $\widetilde{\mathsf{T}}$ as the set of elements of $\mathsf{T}$ for which line S3 was executed. Since Lemma 8 is still true with the same proof, we have $\mathsf{E}\left[|\widetilde{\mathsf{T}}|\right] \leq \log_2 m + \frac{1}{2}$ for all $m$.

We will distribute the average cost of the obtained solution as follows. Each element of $\widetilde{\mathsf{T}}$ gives a charge to $\mathsf{T}$ and a charge to its elements. If the algorithm have received the set of element $X \subseteq U$, then clearly $|\mathcal{T}^*| \geq \frac{|X| \cdot k}{d}$; our goal is to give charges to the elements so that their expected sum equals $xk/d \leq |\mathcal{T}^*|$.

We will again perform an analysis of the average cost of receiving an element $i$ for which the test in line A2 is false. We define or redefine the following notations:

    $\sigma\alpha p[i] = \sum_{S \in \mathcal{S}_i - \mathcal{T}^*} \alpha p[S]$;
    $\xi(i)$ is the value of $\sigma\alpha p[i]$ when line A1 is executed for $i$;
    $\beta(i) = |(\mathcal{S}_i \cap \mathcal{T}^*) - (\mathcal{S}_i \cap \mathcal{T})|$;
    $\psi(i) = |(\mathcal{S}_i \cap \mathcal{T}) - (\mathcal{S}_i \cap \mathcal{T}^*)|$;

The value of deficit in line A1 is at most $\beta(i) - \psi(i)$. Element $i$ will belong to some $\widetilde{T}$ only if $\psi(i) < \beta(i)$. We will view $\xi(i)$ and $\beta(i)$ as fixed parameters of the event when $i$ is received. The quantity $\psi(i)$ is the number of successes in independent trials with success probabilities that add to $\xi(i)$. Let $p(i) = \Pr[\psi(i) < \beta(i)]$.

We charge element $i$ with a value of $\pi_e(i) = \frac{k}{dp(i)}$. The intuition is that, because we make this charge with probability $p(i)$, on an average it equals $p(i)\pi_e(i) = k/d$ and the sum of these charges therefore cannot be larger that $|\mathcal{T}^*|$. We then distribute the remaining cost equally among $\psi(i) < \beta(i)$ many elements of $(\mathcal{S}_i \cap \mathcal{T}^*) - (\mathcal{S}_i \cap \mathcal{T})$.

Clearly, each of the value of deficit computed in line A1 and computed in line A10 cannot exceed $\beta(i)$. The term $\text{deficit} \cdot |\mathcal{S}_i|^{-1}$ in line A6' adds at most deficit to the sum of probabilities computed in line A6', thus the cost attributable to this term, as well as the cost due to line A12 add to at most 2 per $T \in \mathcal{T}^*$. It remains to estimate the cost due to the terms $\alpha p[S]$. We decrease this cost by the charge made to $i$, so each set $T \in \mathcal{T}^*$ such that $i \in \widetilde{T}$ receives a charge of at most $\pi_s(i) = \max\left\{0, \frac{\xi(i) - \pi_e(i)}{\beta(i)}\right\} = \max\left\{0, \frac{\xi(i) - \frac{k}{dp(i)}}{\beta(i)}\right\}$.

The expected number of sets selected by us is therefore at most

$$\begin{aligned}
&\sum_{T \in \mathcal{T}^*} \sum_{i \in \widetilde{T}} (\pi_s(i) + 2) + \sum_{i \in X} p(i)\pi_e(i) \\
\leq\ & |\mathcal{T}^*| \cdot \sum_{i \in \widetilde{T}} \pi_s(i) + 2 \cdot |\widetilde{T}| \cdot |\mathcal{T}^*| + \frac{|X| \cdot k}{d} \\
\leq\ & \left(\left(\tfrac{1}{2} + \log_2 m\right) \pi_s(i) + 2\log_2 m + 1\right) \cdot |\mathcal{T}^*|
\end{aligned}$$

which means we need to estimate the quantity $\pi_s(i)$. For this, we first need to calculate a bound for $p(i)$. Remember that $\psi(i)$ is the number of successes of a set of independent trials with success probabilities that add up to $\xi(i)$. The standard Chernoff bound theorem [9, 16] states that if we have a set of independent trials with the sum of success probabilities $\mu$, the probability that the number of successes is below $(1-\delta)\mu$ is below $e^{-\delta^2 \mu/2}$. In our case, $\mu = \xi(i)$ and $(1-\delta)\mu$ is $\beta(i)$. We introduce the following notations for simplicity: $\beta = \beta(i)$, $\phi = \xi(i)/\beta$ and $\kappa = d/k$. Now $\mu = \phi\beta$ and $\delta = (\phi-1)/\phi$; thus via Chernoff bound we have $p(i) < e^{-\frac{(\phi-1)^2}{2\phi^2}\phi\beta} = e^{-\frac{(\phi-1)^2}{2\phi}\beta}$. Hence

$$\pi_s(i) < \max\left\{0, \phi - \frac{1}{\kappa\beta} e^{\frac{(\phi-1)^2}{2\phi}\beta}\right\} < \max\left\{0, \phi - \frac{1}{\kappa\beta} e^{(\frac{\phi}{2}-1)\beta}\right\}$$

By using simple calculus and the fact that $\beta \geq 1$, it can be shown that the maximum value of the function $f(\phi) = \phi - \frac{1}{\kappa\beta} e^{(\frac{\phi}{2}-1)\beta}$ is at most $2\ln\kappa + 2\ln(2e) < 2\ln\kappa + 3.4$. This shows that

$$\pi_s(i) < \begin{cases} 2\ln\kappa + 3.4 & \text{if } k < (2e) \cdot d \\ 0 & \text{otherwise} \end{cases}$$

### 4.3 Lower bounds on competitive ratios for $\mathbf{OSC}_k$ and $\mathbf{WOSC}_k$

**Lemma 11** [4] *There exists an instance with $m = |\mathcal{S}|$ sets over $n = |V|$ elements such that for any fixed $\delta > 0$ any deterministic algorithm must have a competitive ratio of*

**(i)** $\Omega\left(\frac{\log \frac{m}{k} \log \frac{n}{k}}{\log\log \frac{m}{k} + \log\log \frac{n}{k}}\right)$ *for* $\mathbf{OSC}_k$ *provided* $k \log_2 \frac{n}{k+1} \leq m \leq (k+1)e^{\left(\frac{n}{k}\right)^{\frac{1}{2}-\delta}}$ *and* $k < \min\{m, n\}$;

**(ii)** $\Omega\left(\frac{\log m \log n}{\log\log m + \log\log n}\right)$ *for* $\mathbf{WOSC}_k$ *provided*

$$k + \log_2\left(n - 1 - \lceil \log_2(k+1) \rceil\right) \leq m \leq k + e^{(n-1-\lceil\log_2(k+1)\rceil)^{\frac{1}{2}-\delta}} \text{ and } k < \tfrac{1}{2} \cdot \min\{m, 2^{n-1}\}.$$

---

[4]The relationships between $m$, $n$ and $k$ were referred to as "for almost all values of the parameters" before.

**Proof.**
**(i)** Alon *et al.* [2] provided an instance of **OSC** with $m'$ sets and $n'$ elements with such that the optimal (offline) cover contains just one set but any online cover must use $\Omega\left(\frac{\log m' \log n'}{\log\log m' + \log\log n'}\right)$ sets as long as $\log_2 n' \leq m' \leq e^{(n')^{\frac{1}{2}-\delta}}$ for any fixed $\delta > 0$. Consider a given $k$. We will use one additional element $x$ and $k$ additional sets such that $x$ appears in all these sets. To make these $k$ sets mutually different, we will use an additional $\lceil \log_2(k+1) \rceil$ elements (which we will never present) and add a distinct subset of these additional elements to each of the $k$ sets. We will also have $k$ copies of the instances of Alon *et al.* [2] with elements renamed to make each copy distinct from the rest; each element of each copy is also added to exactly $k-1$ of the $k$ additional sets we mentioned at first. The total number of elements $n$ satisfies $kn' < n = kn' + 1 + \lceil \log_2 k \rceil < (k+1)n'$, and the total number of sets is $m = k + km' < (k+1)m'$ since $k < m$. We first present the element $x$ to force the adversary to select the $k$ additional sets; these sets also cover any element in the $k$ copies of Alon *et al.* [2] exactly $k-1$ times. After this, we present the elements in the $k$ copies of Alon *et al.* [2] following their scheme, presenting elements in one copy completely before presenting elements in the next copy. Now the optimal uses at most $2k$ sets, whereas by a reasoning similar to that in Alon *et al.* [2] any online algorithm must use $\Omega\left(k + k \cdot \frac{\log m' \log n'}{\log\log m' + \log\log n'}\right)$ sets; thus the performance ratio is at least $\Omega\left(\frac{\log m' \log n'}{\log\log m' + \log\log n'}\right) = \Omega\left(\frac{\log \frac{m}{k} \log \frac{n}{k}}{\log\log \frac{m}{k} + \log\log \frac{n}{k}}\right)$. Moreover, the relationship between $m$ and $n$ is given by

$$k \cdot \log_2 \frac{n}{k+1} < k \cdot \log_2 n' \leq km' < m < (k+1)m' \leq (k+1) \cdot e^{(n')^{\frac{1}{2}-\delta}} < (k+1) \cdot e^{\left(\frac{n}{k}\right)^{\frac{1}{2}-\delta}}$$

**(ii)** We again use one additional element $x$ plus $\lceil \log_2(k+1) \rceil$ additional elements (that we will never present) to create $k$ additional sets such that $x$ appears in all these sets. We set the cost of each of these sets to be arbitrarily close to zero, say $\varepsilon$. This time we just use one copy of the instance of Alon *et al.* [2] with each set of cost 1 and, as before, each element of this copy is also added to exactly $k-1$ of the $k$ additional sets we mentioned at first. The total number of elements $n$ satisfies $n' < n = n' + 1 + \lceil \log_2 k \rceil$, and the total number of sets $m$ satisfies $m' < m = k + m'$. We again first present the element $x$ to force the adversary to select the $k$ additional sets; these sets also cover any element in the copy of Alon *et al.* [2] exactly $k-1$ times. After this, we present the elements in the copy of Alon *et al.* [2] with $n'$ elements and $m'$ sets following their scheme. Overall, the optimal uses sets of total cost $1 + \varepsilon$ whereas by a reasoning similar to that in Alon *et al.* [2] any online algorithm must use sets of total cost at least $\varepsilon + \Omega\left(\frac{\log m' \log n'}{\log\log m' + \log\log n'}\right)$; thus setting $\varepsilon$ to be sufficiently small we achieve a competitive ratio of

$$\begin{aligned}
& \Omega\left(\frac{\log m' \log n'}{\log\log m' + \log\log n'}\right) \\
=\ & \Omega\left(\frac{\log(m-k)\log(n-1-\lceil\log_2(k+1)\rceil)}{\log\log(m-k)+\log\log(n-1-\lceil\log_2(k+1)\rceil)}\right) \\
=\ & \Omega\left(\frac{\log m \log n}{\log\log m + \log\log n}\right)
\end{aligned}$$

where the last equality holds since $k < \frac{1}{2} \cdot \min\{m, 2^{n-1}\}$. Moreover, the relationship between $m$ and $n$ is given by

$$k + \log_2(n - 1 - \lceil \log_2(k+1) \rceil) = k + \log_2 n' \leq k + m' = m \leq k + e^{(n')^{\frac{1}{2}-\delta}} = k + e^{(n-1-\lceil \log_2(k+1)\rceil)^{\frac{1}{2}-\delta}}$$

□

**Acknowledgments.** DasGupta thanks the organizers of the Online Algorithms 2004 Workshop (OLA-2004) in Denmark for invitation which provided motivations to look at these cover problems. We also thank Eduardo Sontag for providing us with valuable insights into the biological applications of the online problems, and Yossi Azar for sending us their recent results [1, 3] as well as very useful discussions that led to better understanding of these results.

# References

[1] N. Alon, B. Awerbuch, Y. Azar, N. Buchbinder and J. Naor. *A general approach to online network optimization problems,* proceedings of the 15th ACM-SIAM Symposium on Discrete Algorithms, pp. 570-579, 2004.

[2] N. Alon, B. Awerbuch, Y. Azar, N. Buchbinder, and J. Naor. *The online set cover problem*, proceedings of the 35th annual ACM Symposium on the Theory of Computing, pp. 100-105, 2003.

[3] N. Alon, Y. Azar and S. Gutner. *Admission control to minimize rejections and online set cover with repetitions*, proceedings of the 17th ACM Symposium on Parallelism in Algorithms and Architectures, Las Vegas, NV, USA, July 17-20, pp. 238-244, 2005.

[4] M. Andrec, B.N. Kholodenko, R.M. Levy, and E.D. Sontag. *Inference of signaling and gene regulatory networks by steady-state perturbation experiments: Structure and accuracy*, J. Theoretical Biology, Vol. 232, No. 3, pp. 427-441, 2005.

[5] B. Awerbuch, Y. Azar, A. Fiat and T. Leighton. *Making commitments in the face of uncertainty: how to pick a winner almost every time*, proceedings of the 28th annual ACM Symposium on the Theory of Computing, pp. 519-530, 1996.

[6] P. Berman, B. DasGupta and M. Kao. *Tight approximability results for test set problems in bioinformatics*, Journal of Computer & Systems Sciences, Vol. 71, Issue 2, pp. 145-162, 2005.

[7] P. Berman, B. DasGupta and E. Sontag. *Randomized approximation algorithms for set multi-cover problems with applications to reverse engineering of protein and gene networks*, Discrete Applied Mathematics, Vol. 155, Issues 6-7, pp. 733-749, 2007.

[8] P. Berman, B. DasGupta and E. Sontag. *Algorithmic issues in reverse engineering of protein and gene networks via the modular response analysis method*, to appear in Annals of the New York Academy of Sciences (volume title: Reverse Engineering Biological Networks: Opportunities and Challenges in Computational Methods for Pathway Inference, edited by Gustavo Stolovitsky, Andrea Califano and Jim Collins), November 2007.

[9] H. Chernoff. *A measure of asymptotic efficiency of tests of a hypothesis based on the sum of observations*, Annals of Mathematical Statistics, 23: 493–509, 1952.

[10] E.J. Crampin, S. Schnell, and P.E. McSharry. *Mathematical and computational techniques to deduce complex biochemical reaction mechanisms*, Progress in Biophysics & Molecular Biology, 86, pp. 77-112, 2004.

[11] U. Feige. *A threshold for approximating set cover*, Journal of the ACM, Vol. 45, 1998, pp. 634-652.

[12] D. S. Johnson. *Approximation algorithms for combinatorial problems*, Journal of Computer and Systems Sciences, Vol. 9, 1974, pp. 256-278.

[13] B. N. Kholodenko, A. Kiyatkin, F. Bruggeman, E.D. Sontag, H. Westerhoff, and J. Hoek. *Untangling the wires: a novel strategy to trace functional interactions in signaling and gene networks*, Proceedings of the National Academy of Sciences USA 99, pp. 12841-12846, 2002.

[14] B. N. Kholodenko and E.D. Sontag. *Determination of functional network structure from local parameter dependence data*, arXiv physics/0205003, May 2002.

[15] N. Littlestone. *Learning quickly when irrelevant attributes abound: a new linear-threshold algorithm*, Machine Learning, 2, pp. 285-318, 1988.

[16] R. Motwani and P. Raghavan. *Randomized algorithms*, Cambridge University Press, New York, NY, 1995.

[17] P. Slavik. *A tight analysis of the greedy algorithm for set cover*, proceedings of the 28th ACM Symposium on Theory of Computing, 1996, pp. 435-439.

[18] E.D. Sontag, A. Kiyatkin, and B.N. Kholodenko. *Inferring dynamic architecture of cellular networks using time series of gene expression, protein and metabolite data*, Bioinformatics 20, pp. 1877-1886, 2004.

[19] A. Srinivasan. *Improved approximations of packing and covering problems*, proceedings of the 27th Annual ACM Symposium on Theory of Computing, 1995, pp. 268-276.

[20] J. Stark, R. Callard and M. Hubank. *From the top down: towards a predictive biology of signaling networks*, Trends Biotechnol. 21, pp. 290-293, 2003.

[21] L. Trevisan. *Non-approximability results for optimization problems on bounded degree instances*, proceedings of the 33rd annual ACM Symposium on the Theory of Computing, pp. 453-461, 2001.

[22] V. Vazirani. *Approximation algorithms*, Springer-Verlag, July 2001.

## Appendix

## A  Some combinatorial and probabilistic facts and results

**Fact 1** *If $f$ is a non-negative integer random function, then $E[f] = \sum_{i=1}^{\infty} \Pr[f \geq i]$.*

**Fact 2** *The function $f(x) = xe^{-x}$ is maximized for $x = 1$.*

The subsequent lemmas deal with $N$ independent 0-1 random variables $\tau_1, \ldots, \tau_N$ called trials with event$\{\tau_i = 1\}$ is the *success* of trial number $i$ and $s = \sum_{i=1}^{N} \tau_i$ is the number of successful trials. Let $x_i = \Pr[\tau_i = 1] = E[\tau_i]$ and $X = \sum_{i=1}^{N} x_i = E[s]$.

**Lemma 12** *If $0 < 2\alpha \leq X + 1$ than $\Pr[s = \alpha] > \Pr[s = \alpha - 1]$.*

**Proof.** Our elementary events are 0/1 vectors $\tau = (\tau_1, \ldots, \tau_N)$. Let $E_\alpha$ be the event $\{s = \alpha\}$, *i.e.* the set of elementary events with $\alpha$ 1's. Given $\tau \in E_{\alpha-1}$ we can form an elementary event from $E_\alpha$ by converting some 0 into 1. If we do it with $\tau_i$, call the result $\tau^i$; observe that $\Pr[\tau^i] > x_i \Pr[\tau]$. Therefore the sum of probabilities of elementary events formed from $\tau$ is at least $\Pr[\tau] \sum_{i: \tau_i=0} x_i \geq (X - \alpha + 1)\Pr[\tau] \geq \alpha \Pr[\tau]$.

This shows that the sum of probabilities of the multi-set of elementary events formed from elements of $E_{\alpha-1}$ is larger than $\alpha \Pr[E_{\alpha-1}]$; in turn, every elements in this multi-set belongs to $E_\alpha$ and it is present in this multi-set exactly $\alpha$ times. Thus $\Pr[E_\alpha] \geq \alpha^{-1}\alpha\Pr[E_{\alpha-1}]$. ❏

**Lemma 13** *If $0 \leq \alpha \leq X/2$ then $\Pr[s \leq \alpha] < e^{-X}X^\alpha/\alpha!$.*

**Proof.** The case of $\alpha = 0$ is easy since $\Pr[s \leq 0] = \prod_{i=1}^n (1 - x_i) < \prod_{i=1}^n e^{-x_i} = e^{-X}$. So, we assume in the remaining that $\alpha > 0$.

We will show how to alter the probabilities so that $X$ remains constant and $\Pr[s \leq \alpha]$ does not decrease. Let $x_0 = x_1 + x_2$, $s' = s - \tau_1 - \tau_2$ and let $q_\alpha = \Pr[s' \leq \alpha]$. We assume that $x_0 \leq 1$. Then

$$\begin{aligned}
\Pr[s \leq \alpha] &= \Pr[\tau_1 = \tau_2 = 0 \ \& \ s' \leq \alpha] + \Pr[\tau_1 + \tau_2 = 1 \ \& \ s' \leq \alpha - 1] \\
&\quad + \Pr[\tau_1 = \tau_2 = 1 \ \& \ s' \leq \alpha - 2] \\
&= (1-x_1)(1-x_1)q_\alpha + [(1-x_1)x_2 + x_1(1-x_2)]q_{\alpha-1} + x_1 x_2 q_{\alpha-2} \\
&= (1 - x_0 + x_1 x_2)q_\alpha + (x_0 - 2x_1 x_2)q_{\alpha-1} + x_1 x_2 q_{\alpha-2} \\
&= [P = (1-x_0)q_\alpha + x_0 q_{\alpha-1}] + x_1 x_2(q_\alpha - 2q_{\alpha-1} + q_{\alpha-2}) \\
&= P + x_1 x_2(\Pr[s' = \alpha] - \Pr[s' = \alpha - 1])
\end{aligned}$$

If we keep $x_1 + x_2$ fixed, $P$ is constant and we maximize the latter expression when $x_1 = x_2$ (because $2\alpha \leq (X - x_1 - x_2) + 1$, by Lemma 12, the difference of probabilities in the parenthesis is positive).

This shows that $\Pr[s = \alpha]$ is maximized when all $x_i$'s are equal. We can "pad" the vector of $x_i$'s with zeros, *i.e.* add trials with zero probability of success. This shows that we can overestimate our probability when we go to the limit with $N \to \infty$ and all $x_i$'s equal to $X/N$. We can now finish the proof by observing the following from standard estimates in probability theory:

$$\lim_{N \to \infty} \frac{N!}{(N-\alpha)!\alpha!}\left(1 - \frac{X}{N}\right)^{N-\alpha}\left(\frac{X}{N}\right)^\alpha = \frac{X^\alpha}{e^X \alpha!}$$

❏